\documentclass{amsart}
\newcommand{\R}{\mathbb{R}}

\usepackage{amsmath}%
\usepackage{amsfonts}%
\usepackage{amssymb}%
\usepackage{graphicx}
\newtheorem{theorem}{Theorem}
\theoremstyle{plain}

\newtheorem{definition}{Definition}

\newtheorem{lemma}{Lemma}

\newtheorem{preremark}{Remark}
\newenvironment{remark}{\begin{preremark}\rm}{\end{preremark}}

\numberwithin{equation}{section}
\begin{document}
\title[Uniqueness Weak Solutions Vlasov-Darwin System]{Uniqueness of the compactly supported weak solutions of the relativistic Vlasov-Darwin system}
\author{Reinel Sospedra-Alfonso}
\author{Martial Agueh}
\address
{Institute of Applied Mathematics,\newline%
\indent University of British Columbia, Mathematics Road, Vancouver BC, Canada V6T 1Z2.}%
\email[R. Sospedra-Alfonso]{sospedra@chem.ubc.ca}%
\address{Department of Mathematics and Statistics,\newline%
\indent University of Victoria, PO BOX 3045 STN CSC, Victoria BC, Canada V8W 3P4.}%
\email[M. Agueh]{agueh@math.uvic.ca}%

\date{August 29, 2012}

\begin{abstract}
We use optimal transportation techniques to show uniqueness of the compactly supported weak solutions of the relativistic Vlasov-Darwin system. Our proof extends the method used by Loeper in \cite{Loeper} to obtain uniqueness results for the Vlasov-Poisson system.
\end{abstract}
\maketitle


\section{Introduction}
\label{Intro}

The relativistic Vlasov-Darwin (RVD) system describes the evolution of a collisionless plasma whose particles interact only through the electromagnetic field they induce. In contrast to the Vlasov-Maxwell system, the particle interaction is assumed to be a low-order relativistic correction (i.e., the Darwin approximation) of the full Maxwell case.

In terms of the electromagnetic field, the RVD system can be introduced as follows. Consider an ensemble of identical charged particles with mass $m$ and charge $q$, which we set to one for simplicity. Let $f(t,x,\xi)$ denote the density of the particles in the phase space $\R^3_x\times\R^3_\xi$ at a time $t\in\left]0,\infty\right[$, where $x$ denotes position and $\xi$ momentum. If collisions are neglected, the time evolution of $f$ is given by the Vlasov equation
\begin{equation}
\label{Vlasov For Maxwell}
\partial_tf+v\cdot\nabla_{x}f+\left(E+c^{-1}v\times B\right)\cdot\nabla_{\xi}f=0, 
\end{equation}
where $v=\frac{\xi}{\sqrt{1+c^{-2}\left|\xi\right|^2}}$ is the relativistic velocity and $c$ is the speed of light.  The electric field $E=E(t,x)$, and magnetic field $B=B(t,x)$, are induced by the particles and satisfy Maxwell's equations
\begin{eqnarray}
\label{Maxwell 1}
\nabla\times B-c^{-1}\partial_tE & = & 4\pi c^{-1}j,\quad\quad \nabla\cdot B\quad  =\quad  0,\\
\label{Maxwell 2}
\nabla\times E +c^{-1}\partial_tB & = & 0, \quad\quad\quad\quad\ \, \nabla\cdot E\quad  =\quad  4\pi\rho,
\end{eqnarray}
where 
\begin{equation}
\label{Density and Current}
  \rho(t,x)=\int_{\mathbb{R}^3}f(t,x,\xi)d\xi \quad \hbox{and} \quad j(t,x)=\int_{\mathbb{R}^3}v(\xi)f(t,x,\xi)d\xi
\end{equation}
are respectively the charge and current densities. The coupled nonlinear system (\ref{Vlasov For Maxwell})-(\ref{Density and Current}) is known as the relativistic Vlasov-Maxwell (RVM) system, which is essential in the study of dilute hot plasmas; see \cite{GlasseyBook} and the references therein. 

The electric field $E$ can be further decomposed into a longitudinal component $E_L$, and a transverse component $E_T$, as
\begin{equation}
\label{Components Electric Field}
E=E_L + E_T,  \quad \nabla\times E_L=0, \quad \nabla\cdot E_T=0.
\end{equation}
The Darwin approximation consists of neglecting the transverse part of the displacement current, $\partial_tE_T$, in Maxwell-Amp\`ere's law (\ref{Maxwell 1}). Maxwell's equations (\ref{Maxwell 1})-(\ref{Maxwell 2}) then reduce to
\begin{eqnarray}
\label{Darwin 1}
\nabla\times B-c^{-1}\partial_tE_L & = & 4\pi c^{-1}j,\quad \nabla\cdot B\quad  =\quad  0,\\
\label{Darwin 2}
\nabla\times E_T +c^{-1}\partial_tB & = & 0, \quad\quad\quad \nabla\cdot E_L\quad =\quad  4\pi\rho.
\end{eqnarray}
The RVD system is defined as the Vlasov equation (\ref{Vlasov For Maxwell}) coupled with (\ref{Darwin 1})-(\ref{Darwin 2}) via the charge and current densities (\ref{Density and Current}). From the physical point of view, this model neglects the high-frequency electromagnetic waves but describes a wide variety of phenomena that take place in the low-frequency range \cite{bms2007}. 

In \cite{ARMA}, we showed that the  RVD system can be written in terms of a scalar potential $\Phi$, a divergence free vector potential $A$ (i.e., $\nabla\cdot A=0$), and the variables $(x,p)$ in the generalized phase space $\R^3_x\times\R^3_p$, as
\begin{equation}
\label{Vlasov In RVD}
\partial_tf+v_A\cdot\nabla_xf-\left[\nabla\Phi-c^{-1}v^i_A\nabla A^i\right]\cdot\nabla_{p}f=0,\quad  v_A=\frac{p-c^{-1}A}{\sqrt{1+c^{-2}\left|p-c^{-1}A\right|^2}},
\end{equation}
coupled with
\begin{eqnarray}
\label{Potentials Phi}
\Phi(t,x) & = & \int_{\mathbb{R}^3}\rho(t,y)\frac{dy}{\left|y-x\right|},\\ 
\label{Potentials A}
A(t,x) & = & \frac{1}{2c}\int_{\mathbb{R}^3}\left[\texttt{id}+\omega\otimes\omega\right]j_A(t,y)\frac{dy}{\left|y-x\right|},
\end{eqnarray}
via
\begin{equation}
\label{Densities}
\quad \rho(t,x)=\int_{\mathbb{R}^3}f(t,x,p)dp, \quad j_A(t,x)=\int_{\mathbb{R}^3}v_Af(t,x,p)dp.
\end{equation}
The one-particle distribution function $f=f(t,x,p)$ depends on time $t\in]0,\infty[$, position $x\in\mathbb{R}^3$ and generalized momentum $p\in\mathbb{R}^3$,  where $p:=\xi+c^{-1}A$. The scalar and vector potentials $\Phi$ and $A$ are induced by $f$ via the charge and current densities $\rho$ and $j_A$, respectively. The relativistic particle velocity is denoted by $v_A$, and satisfies $\left|v_A\right|\leq c$ where $c$ is the speed of light. For the sake of notation, we denote the $3$-by-$3$ identity matrix as $\texttt{id}$, and the unit vector $\omega=\left(y-x\right)/\left|y-x\right|$. The symbol $\otimes$ in (\ref{Potentials A}) stands for the tensor product, i.e., $\omega\otimes\omega$ is the $3$-by-$3$ matrix with entries $\omega^i\omega^j$, where $\omega=(\omega^1, \omega^2,\omega^3)$. As usual, the repeated indexes in (\ref{Vlasov In RVD}) means summation, i.e.,  $v^i_A\nabla A^i \equiv \sum_{i=1}^3 v^i_A\nabla A^i$. 

We briefly comment on the potential formulation (\ref{Vlasov In RVD})-(\ref{Densities}) of the RVD system; for more details we refer to \cite{ARMA}. From the two homogeneous Maxwell's equations in (\ref{Maxwell 1})-(\ref{Maxwell 2}), we can easily show that the electric and magnetic fields $(E,B)$ can be represented in terms of a scalar and vector potentials  $(\Phi,A)$ as
\begin{eqnarray}
\label{Electric Field}
E(t,x) & = & -\nabla\Phi(t,x)-c^{-1}\partial_tA(t,x),\\
\label{Magnetic Field}
B(t,x) & = & \nabla\times A(t,x),
\end{eqnarray}
so that Maxwell's equations (\ref{Maxwell 1})-(\ref{Maxwell 2}) in the Coulomb gauge (i.e., $\nabla\cdot A =0$) read
\begin{eqnarray}
\label{Maxwell potential form 1}
 \Delta\Phi & = &-4\pi\rho,\\
\label{Maxwell potential form 2} 
 \Delta A -c^{-2}\partial^2_{t} A & = & -4\pi c^{-1}j + c^{-1} \nabla(\partial_t \Phi).
\end{eqnarray}
Since by (\ref{Components Electric Field}) and (\ref{Electric Field}) we have $E_T = c^{-1}\partial_tA$, then the Darwin approximation yields
\begin{eqnarray}
\label{Darwin potential form 1}
 \Delta\Phi & = & -4\pi\rho \\
\label{Darwin potential form 2}
 \Delta A & = & -4\pi c^{-1}j + c^{-1} \nabla(\partial_t \Phi).
 \end{eqnarray}
This approximation does not affect the scalar potential in any way, but neglects the retardation effects of the vector potential. In particular, the hyperbolic equation (\ref{Maxwell potential form 2}) is reduced to the more tractable elliptic equation (\ref{Darwin potential form 2}). 

We proved in \cite{ARMA} that the potentials $(\Phi,A)$ defined as in (\ref{Potentials Phi})-(\ref{Potentials A}) solve (\ref{Darwin potential form 1})-(\ref{Darwin potential form 2}). As for the formulation (\ref {Vlasov In RVD}) of the Vlasov equation (\ref{Vlasov For Maxwell}), it is derived via the characteristic system of (\ref{Vlasov For Maxwell}), where we use the change of variables $(x,p):=(x,\xi + c^{-1}A)$. Under this change of variables, it is easy to check that the relativistic velocity $v_A$, the charge density $\rho$, and the current density $j_A$, are of the form given by (\ref{Vlasov In RVD}) and (\ref{Densities}).

\smallskip

\begin{definition}
\label{Definition Solution}
Let $f_0\in L^1\cap L^{\infty}(\R^6;\mathbb{R})$ be non-negative. For $T>0$ consider $f\in C([0,T[,L^{\infty}(\mathbb{R}^6)-w*;\mathbb{R})$ such that $f \geq 0$. We call $f$ a weak solution of the relativistic Vlasov-Darwin system with initial datum $f_0$, if
$$\int_{\R^3}\int_{\R^3}f(t,x,p)\,dxdp=\int_{\R^3}\int_{\R^3} f_0(x,p)\,dxdp \quad \forall t\in [0,T[, $$ 
$f$ induces potentials $\Phi$ and $A$ given by (\ref{Potentials Phi})-(\ref{Densities}), and for all \\  $\varphi\in C^{\infty}_0([0,T[\times\mathbb{R}^6;\mathbb{R})$ we have
\begin{eqnarray*}
\label{Solution Distributions} 
\int^T_0\int_{\R^3}\int_{\R^3}f\left\{\partial_t\varphi+v_A\cdot\nabla_x\varphi-\left[\nabla\Phi-c^{-1}v^i_A\nabla A^i\right]\cdot\nabla_p\varphi\right\}(t,x,p)dtdxdp\\
=-\int_{\R^3}\int_{\R^3}f_0(x,p)\varphi(0,x,p)\,dxdp.
\end{eqnarray*}
\end{definition}

The notation $\varphi\in C^{\infty}_0([0,T[\times\mathbb{R}^6;\mathbb{R})$ means that $\varphi$ is of class $C^\infty$, $\varphi(T)=0$ and $\varphi(t)$ has a compact support in $\R^6$ for all $t\in[0,T[$. Also, $L^{\infty}(\mathbb{R}^6)-w*$ means that the space $L^{\infty}(\mathbb{R}^6)$ is equipped with the weak-* topology.

\smallskip

The global in time existence of weak solutions to the RVD system without size restriction on the Cauchy datum  was established by Pallard in \cite{Pallard}. This result uses the formulation of the RVD system in terms of the electromagnetic field discussed earlier, i.e., (\ref{Vlasov For Maxwell}), (\ref{Density and Current}) and (\ref{Darwin 1})-(\ref{Darwin 2}). Here we do not discuss the existence problem nor the equivalence of the two formulations in the context of weak solutions. We are only concerned with the uniqueness of solutions to (\ref{Vlasov In RVD})-(\ref{Densities}) in the sense of Definition \ref{Definition Solution}. We emphasize, however, that for a compactly supported distribution function, Definition \ref{Definition Solution} makes perfect sense. We postpone this discussion to the next section (see Lemmas \ref{Lemma Properties Potentials} and \ref{Definition Linear Vlasov}).

If we formally let $c\rightarrow\infty$ in (\ref{Vlasov In RVD})-(\ref{Densities}), the model equations reduce to the Vlasov-Poisson system, which is the zeroth-order relativistic correction of the Vlasov-Maxwell system. A uniqueness result for weak solutions of the Vlasov-Poisson system based on optimal transportation, under the assumption that the charge density remains bounded, was established by Loeper in \cite{Loeper}; see also Robert \cite{Robert} for the uniqueness of compactly supported weak solutions. Here, we extend the method of \cite{Loeper} to produce a uniqueness of weak solutions to the RVD system under the assumption of a compactly supported distribution function. To the best of our knowledge, this is the first uniqueness result for weak solutions of the RVD system.  Our main result is the following:

\begin{theorem}  
\label{Main Theorem}
Let $f_0\in L^1\cap L^{\infty}(\R^6;\mathbb{R})$, $f_0\geq 0$, have compact support and let $f$ be a weak solution on $[0,T[$ of the RVD system (\ref{Vlasov In RVD})-(\ref{Densities}) in the sense of Definition \ref{Definition Solution}, such that $\left.f\right|_{t=0}=f_0$. If for all $t\in [0,T[$ the support of $f(t)$ is compact in $\mathbb{R}^6$, then this solution is unique. 
\end{theorem}

The proof of the theorem will be postponed to Section \ref{section proof}. As mentioned earlier, it uses techniques of optimal transportation theory and extends the proof by Loeper in \cite{Loeper} for the uniqueness of solutions to the Vlasov-Poisson system. These techniques have been useful in the recent paper by Carrillo and Rosado \cite{CR} where uniqueness of solutions to several equations containing aggregation terms and aggregation/diffusion competition (e.g. swarming models, chemotaxis) was established. The main new difficulty in our proof is the vector potential $A$, which for a given $f$ is defined by a nonlinear integral equation. The existence and uniqueness of a solution to this integral equation, its regularity, and the corresponding a-priori estimates used in our proof, require some elaborated work.

The rest of the paper is organized as follows. In Section \ref{Preliminaries Section}, we present some preliminary results on the scalar and vector potentials $\Phi$ and $A$. We also discuss results on the linear Vlasov equation (\ref{Vlasov In RVD}) and the associated characteristic system. In Section \ref{transport section}, we recall well-known results of optimal transportation that we use in Section \ref{Estimates on the Field} to establish some crucial estimates on the potentials, which are needed for the proof of Theorem \ref{Main Theorem}. Hereafter, $L^\infty(X;Y)$ will denote the set of $L^\infty$- functions $f: X\to Y$, and $C^\infty_0(X;Y)$ denotes the set of such functions of class $C^\infty$ with compact support in $X$. By  $g\in L^\infty([0,T[, C_b(X); Y)$, we mean that $[0,T[ \ni t\mapsto g(t)\in C_b(X;Y)$ belongs to  $L^\infty([0,T[)$. For a function $\R^3\times\R^3\ni (x,p)\mapsto g(x,p)$, we denote by $\|g\|_{L^\infty_x}$ (resp. $\|g\|_{L^\infty_p}$) the $L^\infty$-norm of $x\mapsto g(x,p)$ (resp. $p\mapsto g(x,p)$), and by $\texttt{supp}\,g$ the support of $g$. Also $\partial_xg$ (resp. $\partial_pg$) denotes the matrix gradient of $g$ with respect to $x$ (resp. $p$). If $D=[d_{ij}]$ is a real matrix,  we denote the $\infty$-matrix norm of $D$ by $|D|=\sup_{i,j}|d_{ij}|$.


\section{Preliminaries}
\label{Preliminaries Section}

Henceforth, we set the speed of light to $c=1$. For simplicity and without loss of generality we shall omit the time dependence throughout this section, unless we specify otherwise. We first recall an estimate that will be used later on. For a proof see \cite[Lemma 2.7]{Pallard}.

\begin{lemma} 
\label{Pallard}
For $m\in\left\{1,2\right\}$ let $r_0=3/(3-m)$ and $r<r_0<s$. Set $\lambda=\left(1-r/r_0\right)/\left(1-r/s\right)$. Then there exists a constant $C=C(m,r,s)>0$ such that for any $\Psi\in L^r\cap L^s(\mathbb{R}^3;\mathbb{R})$ 
$$ \left\|\int_{\mathbb{R}^n}\Psi(y)\frac{dy}{\left|y-\cdot\right|^m}\right\|_{L^{\infty}_x}\leq C(m,r,s)\left\|\Psi\right\|^{1-\lambda}_{L^{r}_x}\left\|\Psi\right\|^{\lambda}_{L^{s}_x}. 
$$
In particular, $C(m,1,\infty)=3\left(4\pi/m\right)^{m/3}/\left(3-m\right)$.
\end{lemma}


Next, we show an alternative formulation of the integral equation (\ref{Potentials A}) which is used later on. The following technical result will be needed. 

\begin{lemma} 
 \label{Lemma Intermediate Step}
 Denote $\omega=(\omega^1,\omega^2,\omega^3)$; $\nabla_x=(\partial_1,\partial_2,\partial_3)$ and let $\delta_{ik}=1$ if $i=k$ else $\delta_{ik}=0$. For any $\varphi\in C_0^{\infty}(\mathbb{R}^3;\mathbb{R})$ and $y\in\mathbb{R}^3$, we have
\begin{equation}
\label{Intermediate Step} \int_{\mathbb{R}^3}\Delta\varphi(x)\left[\delta_{ik}-\omega^i\omega^k\right]\frac{dx}{\left|y-x\right|}=2\int_{\mathbb{R}^3}\partial_k\partial_i\varphi(x)\frac{dx}{\left|y-x\right|}.
\end{equation}
 \end{lemma}
 
{\em Proof}. Clearly, both integrals are well defined. Let the support of $\varphi$ be contained in a ball centered at $y$ with radius $R>0$. Let $0<r<R$. It is easy to check that $\partial_k\omega^i=-\left|y-x\right|^{-1}\left[\delta_{ik}-\omega^i\omega^k\right]$ and $\Delta\omega^i=-2\left|y-x\right|^{-2}\omega^i$ for $\left|y-x\right|>r$. To prove (\ref{Intermediate Step}), we will show that
\begin{equation}
\label{eqlim}
2\int_{\mathbb{R}^3}\partial_k\partial_i\varphi(x)\frac{dx}{\left|y-x\right|}=-\lim_{r\rightarrow0}\int_{\left|y-x\right|\geq r}\Delta\varphi(x)\partial_k\omega^idx.
\end{equation}
Denote the integral on the right-hand side of (\ref{eqlim}) by $I(r)$. In view of the compact support of $\varphi$, we can restrict the domain of integration of $I(r)$ to $r\leq\left|y-x\right|\leq R$. Then, integration by parts in $\partial_k$ and then twice for $\Delta$ yield
\begin{eqnarray*} 
  I(r) & = & -\int_{r\leq\left|y-x\right|\leq R}\partial_k\varphi(x)\Delta\omega^idx+o(r)\\
  & = & 2\int_{r\leq\left|y-x\right|\leq R}\partial_k\varphi(x)\frac{\omega^idx}{\left|y-x\right|^2}+o(r),
\end{eqnarray*}
where $o(r)$ stands for all boundary terms at $\left|x-y\right|=r$. Note that the boundary terms at $\left|y-x\right|=R$ vanish due to the compact support of $\varphi$. It is not difficult to check that $o(r)\rightarrow0$ as $r\rightarrow0$. Then, since we also have $\left|y-x\right|^{-2}\omega^i=\partial_i\left|y-x\right|^{-1}$ for $\left|y-x\right|>r$, another integration by parts and a standard limiting process produce the identity (\ref{Intermediate Step}).
\qed

\smallskip 

\begin{lemma} 
\label{Lemma Vector Potential}
Let $f$ be given as in Definition \ref{Definition Solution}. Define $\rho$ according to (\ref{Densities}) and suppose that $\rho\in L^{\infty}(\mathbb{R}^3;\mathbb{R})$. If $A$ solves the integral equation (\ref{Potentials A}) for the vector potential, then 
$$
\Delta A(x) = -4\pi j_{A}(x)-\nabla\left\{\nabla\cdot\int_{\mathbb{R}^3}j_{A}(y)\frac{dy}{\left|y-x\right|}\right\}
$$
in the sense of distributions, i.e., for any $\varphi\in C_0^{\infty}(\mathbb{R}^3;\mathbb{R})$, we have
\begin{eqnarray}
\label{Distributions}
\lefteqn{-\int_{\mathbb{R}^3}A(x)\Delta\varphi(x)dx}\nonumber\\
& & =4\pi \int_{\mathbb{R}^3}\varphi(x)j_A(x)dx-\int_{\mathbb{R}^3}\nabla\varphi(x)\left\{\nabla\cdot\int_{\mathbb{R}^3}j_{A}(y)\frac{dy}{\left|y-x\right|}\right\}dx.
\end{eqnarray}
\end{lemma}

{\em Proof}. Since $\left|v_A\right|\leq1$, we have $\left|j_A\right|\leq\rho$ and thus $j_A\in L^{\infty}(\mathbb{R}^3;\mathbb{R}^3)$. Then, the integral in curly brackets on the right-hand side of (\ref{Distributions}) is of class $C^1(\mathbb{R}^3;\mathbb{R}^3)$. Therefore, the last term in (\ref{Distributions}) is well defined and so are the other terms of this equation. Now, substitution of (\ref{Potentials A}) into the left-hand side of (\ref{Distributions}) yields,
\begin{eqnarray}
\label{LHS}
\lefteqn{-\int_{\mathbb{R}^3}A(x)\Delta\varphi(x)dx}\nonumber\\
& = & -\int_{\mathbb{R}^3}\Delta\varphi(x)\left\{\frac{1}{2}\int_{\mathbb{R}^3}\left[\texttt{id}+\omega\otimes\omega\right]j_A(y)\frac{dy}{\left|y-x\right|}\right\}dx\nonumber\\
& = & -\int_{\mathbb{R}^3}\Delta\varphi(x)\left\{\int_{\mathbb{R}^3}\left[\texttt{id}-\frac{1}{2}\left(\texttt{id}-\omega\otimes\omega\right)\right]j_A(y)\frac{dy}{\left|y-x\right|}\right\}dx\nonumber\\
& = & I_1 + I_2,
\end{eqnarray}
where,
\begin{equation}
\label{I1}
I_1 =  -\int_{\mathbb{R}^3}\Delta\varphi(x)\left\{\int_{\mathbb{R}^3}j_A(y)\frac{dy}{\left|y-x\right|}\right\}dx=4\pi\int_{\mathbb{R}^3}\varphi(x)j_A(x)dx.
\end{equation}
In the last step we have used that the integral in curly brackets in (\ref{I1}) is a solution of the Poisson equation $\Delta u=-4\pi j_A$ in the sense of distributions. As for the integral $I_2$, we work per components:
\begin{eqnarray}
\label{I2}
I^i_2 & = &  \frac{1}{2}\int_{\mathbb{R}^3}\Delta\varphi(x)\left\{\int_{\mathbb{R}^3}\left[\delta_{ik}-\omega^i\omega^k\right]j^k_A(y)\frac{dy}{\left|y-x\right|}\right\}dx\nonumber\\
& = & \frac{1}{2}\int_{\mathbb{R}^3}j^k_A(y)\left\{\int_{\mathbb{R}^3}\Delta\varphi(x)\left[\delta_{ik}-\omega^i\omega^k\right]\frac{dx}{\left|y-x\right|}\right\}dy\nonumber\\
& = & \int_{\mathbb{R}^3}j^k_A(y)\left\{\int_{\mathbb{R}^3}\partial_k\partial_i\varphi(x)\frac{dx}{\left|y-x\right|}\right\}dy\nonumber\\
& = & -\int_{\mathbb{R}^3}\partial_i\varphi(x)\left\{\partial_k\int_{\mathbb{R}^3}j^k_A(y)\frac{dy}{\left|y-x\right|}\right\}dx,
\end{eqnarray}
where in the second and third steps we have used Fubini's theorem and Lemma \ref{Lemma Intermediate Step}, respectively, and Fubini's theorem and integration by parts in the last step. Note that the resulting boundary term vanishes in view of the compact support of $\varphi$. The relation (\ref{Distributions}) readily follows from (\ref{LHS})-(\ref{I2}). 
\qed

\smallskip

We now show that the potentials $\Phi$ and $A$ induced by a compactly supported weak solution of the RVD system are well-defined and satisfy the appropriate regularity.

\begin{lemma} 
\label{Lemma Properties Potentials}
Let $f$ be given as in Definition \ref{Definition Solution} and suppose that $f$ has compact support. Then, there exists unique bounded potentials $\Phi$ and $A$ that solve (\ref{Potentials Phi})-(\ref{Densities}). Moreover, $\left(\partial_x\Phi,\partial_xA\right)\in C_b(\mathbb{R}^3;\mathbb{R}\times\mathbb{R}^3)$ and therefore $f\nabla\Phi$ and $fv^i_A\nabla A^i$ are in $L^1(\mathbb{R}^6;\mathbb{R}^3)$, with $v_A$ defined by (\ref{Vlasov In RVD}). In addition, $\nabla\cdot A=\texttt{Trace}(\partial_xA)=0$ where 
$$ \partial_xA(x)=\frac{1}{2}\int_{\mathbb{R}^3}\int_{\mathbb{R}^3}\left\{\omega\otimes v_{A}-v_{A}\otimes\omega+\left[3\omega\otimes\omega-\texttt{id}\right]\left(v_A\cdot\omega\right)\right\}f(y,p)\frac{dpdy}{\left|y-x\right|^2}
$$ 
is the matrix gradient of $A$.
\end{lemma}

{\em Proof}. The results corresponding to the scalar potential $\Phi$ are well-known, see for instance \cite[Chapter 10]{Lieb}; therefore we center our attention on the vector potential $A$.  We prove that the integral equation (\ref{Potentials A}), i.e.,
\begin{equation}
\label{Integral Equation}
A(x)=\frac{1}{2}\int_{\mathbb{R}^3}\int_{\mathbb{R}^3}\left[\texttt{id}+\omega\otimes\omega\right]v_{A}f(y,p)\frac{dpdy}{\left|y-x\right|},\quad v_A=\frac{p-A}{\sqrt{1+\left|p-A\right|^2}},
\end{equation}
has a unique solution which satisfies the claimed regularity. 

\textbf{Uniqueness:}  For $f$ as given, let $A_1,A_2\in C^1_b(\mathbb{R}^3;\mathbb{R}^3)$ be two solutions of (\ref{Integral Equation}). Define $\rho$ as in (\ref{Densities}) and suppose that $f$ has a support included in $B_R\times B_R$, where $B_R$ is a ball of radius $R>0$. Then $\rho$ has support in $B_R$. For $i\in\left\{1,2\right\}$, set $g_{A_i}(x,p)= p-A_i(x)$ and
\[j_{A_i}(x)=\int_{\R^3} v_{A_i}(x,p)f(x,p)\,\mbox{d}p,\quad v_{A_i}=\frac{g_{A_i}}{\sqrt{1+\left|g_{A_i}\right|^2}}.\]
Since $\left|v_{A_i}\right|\leq1$, then $\left|j_{A_i}\right|\leq\rho$ . By the assumptions on $f$, we have that $j_{A_i}\in L^1\cap L^\infty(\R^3;\R^3)$ has support in $B_R$. Then, $j_{A_i}\in L^2(\R^3;\R^3)$, and by the generalized theory of Poisson's equation  \cite[chap. 8]{Trudinger},  the Newtonian potential
$$\int_{\R^3}j_{A_i}(y)\frac{dy}{\left|y-x\right|} \in W^{1,2}(\R^3;\R^3),$$ 
which implies that  
$$\nabla\left\{\nabla\cdot\int_{\mathbb{R}^3}j_{A_i}(y)\frac{dy}{\left|y-x\right|}\right\} \in H^{-1}(\R^3;\R^3).$$
In view of Lemma \ref{Lemma Vector Potential},
 \begin{equation}\label{equation delta Ai}
\Delta A_i(x) = -4\pi j_{A_i}(x)- \nabla\left\{\nabla\cdot\int_{\mathbb{R}^3}j_{A_i}(y)\frac{dy}{\left|y-x\right|}\right\}, \quad i=1, 2,
\end{equation}
in the sense of distributions, and since the right-hand side of (\ref{equation delta Ai}) belongs to $H^{-1}(\R^3;\R^3)$, we deduce by the generalized theory of Poisson's equation that $A_i\in W^{1,2}(\R^3;\R^3)$. Taking the difference of the two equations in (\ref{equation delta Ai}), we obtain
\[-\Delta (A_1-A_2)(x) = 4\pi \left(j_{A_1}-j_{A_2}\right)(x) + \nabla\left\{\nabla\cdot\int_{\mathbb{R}^3} \left(j_{A_1}-j_{A_2}\right)(y)\frac{dy}{\left|y-x\right|}\right\},\]
and since $A_1-A_2\in W^{1,2}(\R^3)$, then integration by parts against $A_1-A_2$ yields
\begin{eqnarray*}
\int_{\R^3} \left|\partial_xA_1 - \partial_xA_2\right|^2\,\mbox{d}x  &=& 4\pi \int_{\R^3} (A_1-A_2)\cdot (j_{A_1}-j_{A_2})\,\mbox{d}x \\
& & \quad - \int_{\R^3}\nabla\cdot(A_1-A_2)\nabla\cdot I\,\mbox{d}x
\end{eqnarray*}
where 
\[ I(x) = \int_{\mathbb{R}^3} \left(j_{A_1}-j_{A_2}\right)(y)\frac{dy}{\left|y-x\right|}.\]
Notice that the boundary terms vanish. Indeed, standard arguments show that both $\partial_xA_i(x)$ and $\partial_xI(x)$ have a decay $O(\left|x\right|^{-2})$, and $A_i(x)$ has a decay $O(\left|x\right|^{-1})$. Then, the products $A_i\partial_xI(x)$ and $A_i\partial_xA_i(x)$ decay like $O(\left|x\right|^{-3})$, which is sufficient for the disappearance of the boundary terms. By Lemma \ref{Lemma Properties Potentials}, $\nabla\cdot A_i=0$. Thus the last integral on the right-hand side vanishes as well. Therefore, 
\begin{equation}
\label{needed}
\int_{\R^3} \left|\partial_xA_1 - \partial_xA_2\right|^2\,\mbox{d}x - 4\pi \int_{\R^3}\int_{\R^3} f(A_1-A_2)\cdot (v_{A_1}-v_{A_2})\,\mbox{d}p\,\mbox{d}x=0.
\end{equation}

We now look for a suitable lower bound on the left-hand side of this equation. To this end, consider the vector-valued function 
\[ \R^3 \ni z \mapsto v(z) = \frac{z}{\sqrt{1+|z|^2}} \in \R^3. \]
Clearly, $v\in C^1(\R^3; \R^3)$ and recall that $v_{A_i}=v(g_{A_i})$ where $g_{A_i}(x,p)=p-A_i(x)$ and $(x,p)\in B_R\times B_R$. It is easy to check that the derivative $Dv(z)$ of $v$ at any point $z\in\R^3$ is given by the 3-by-3 real symmetric matrix 
\[Dv(z)=\frac{1}{\sqrt{1+|z|^2}}\left[\texttt{id}-\frac{z\otimes z}{1+|z|^2}\right]. \]
Then $Dv(z)$ is real orthogonally diagonalizable \cite[Theorem 2.5.6]{Horn}. Moreover, if $z=(z_1, z_2, z_3)\in \R^3$, then $\texttt{det}Dv(z) =(1+|z|^2)^{-5/2} >0$, the determinant of the 1-by-1 upper-left square submatrix of $Dv(z)$ is $(1+z_2^2+z_3^3)(1+|z|^2)^{-3/2} > (1+|z|^2)^{-3/2}>0$, and that of its 2-by-2 upper-left square submatrix is $(1+z_3^3)(1+|z|^2)^{-2} > (1+|z|^2)^{-2}>0$. Therefore the matrix  $Dv(z)$ is  positive definite \cite[Theorem 7.2.5]{Horn}, and we have
\begin{equation}\label{bilinear}
Dv(z) \xi \cdot \xi \geq \lambda|\xi|^2 \qquad \forall (z,\xi)\in \R^3\times\R^3,
\end{equation}
where $\lambda>0$ can be chosen as the lowest eigenvalue of $Dv(z)$. By the mean value theorem, we have at every point $(x,p)\in B_R\times B_R$,
\begin{eqnarray*}
(A_1-A_2)\cdot (v_{A_1}-v_{A_2})  &=&  \left[v(g_{A_1}) - v(g_{A_2})\right] \cdot (A_1-A_2) \\
&=&  Dv\left(g_{A_2} +\delta(g_{A_1}-g_{A_2})\right)\left(g_{A_1}-g_{A_2}\right) \cdot (A_1-A_2)\\
&=& - Dv\left(g_{A_2} +\delta(g_{A_1}-g_{A_2})\right)(A_1-A_2) \cdot (A_1-A_2)
\end{eqnarray*}
for some $\delta\in (0,1)$. But the boundedness of the potentials implies that for some $K_R>0$, the estimate $\left|g_{A_i}(x,p)\right|\leq K_R$ holds for all $(x,p)\in B_R\times B_R$. Then $\left|\left(g_{A_2} +\delta(g_{A_1}-g_{A_2})\right)(x,p)\right| \leq K_R$ uniformly on $(x,p)\in  B_R\times B_R$. Hence, using (\ref{bilinear}) and the above identity,  we have for some constant $C_R>0$ that
\[(A_1-A_2)\cdot (v_{A_1}-v_{A_2}) \leq - C_R |A_1 - A_2|^2\]
uniformly on $B_R\times B_R$. As a result we find from (\ref{needed}) that 
$$
\int_{\R^3} \left|\partial_xA_1 - \partial_xA_2\right|^2\,\mbox{d}x + C_R \int_{\R^3}\int_{\R^3} \rho\left|A_1-A_2\right|^2\mbox{d}x\leq0.
$$
Since the left-hand side is non-negative, and in view of the regularity of the potentials, we deduce that $A_1=A_2$ and therefore uniqueness.

\smallskip

\textbf{Existence:} Let $\bar{C}>0$ be a constant that may depend on $f$, to be fixed later on. Define the set
$$ \mathcal{D}_{\bar{C}}=\left\{A\in C_b(\mathbb{R}^3;\mathbb{R}^3):\left\|A\right\|_{L^{\infty}_x}\leq \bar{C}\right\}. 
$$
We show that there exists some $A_{\infty}\in \mathcal{D}_{\bar{C}}$ which solves (\ref{Integral Equation}). To start with, denote the kernel $\mathcal{K}(x,y)=\left|y-x\right|^{-1}\left[\texttt{id}+\omega\otimes\omega\right]$ and let $A\in \mathcal{D}_{\bar{C}}$.  Since all the entries of the matrix $\left[\texttt{id}+\omega\otimes\omega\right]$ are bounded as $|\delta_{ij} +\omega^i \omega^j|\leq 2$, then the $\infty$-matrix norm of $\mathcal{K}(x,y)$ satisfies $\left|\mathcal{K}(x,y)\right|\leq 2\left|y-x\right|^{-1}$. Consider now the mapping $A\mapsto T[A]$ defined by 
$$
T[A](x)= \frac{1}{2}\int_{\mathbb{R}^3}\int_{\mathbb{R}^3}\mathcal{K}(x,y)v_Af(y,p)dpdy, \quad v_A=\frac{p-A}{\sqrt{1+\left|p-A\right|^2}}. 
$$

We will show that $T$ has a fixed point $A_\infty$ in $ \mathcal{D}_{\bar{C}}$. We claim that $T[A]\in\mathcal{D}_{\bar{C}}$. Indeed, let $\mathcal{K}_{ij}(x,y)$ be the $ij$-entry of $\mathcal{K}(x,y)$, i.e., 
\[\mathcal{K}_{ij}(x,y) = \left(\delta_{ij} + \omega^i_x \omega^j_x\right) \frac{1}{|y-x|},\]
where
\[\omega_x = \frac{y-x}{|y-x|}, \quad \omega_x = (\omega_x^1, \omega_x^2, \omega_x^3).\]

We first show that for $x, y, z\in \R^3$, there exist some $u_1$, $u_2$ and $u_3$ on the line segment between $x$ and $z$, such that 
\begin{eqnarray}
\lefteqn{\left|\mathcal{K}_{ij}(x,y)-\mathcal{K}_{ij}(z,y)\right|} \nonumber\\ 
& \leq & 2\left|\frac{1}{\left|y-x\right|}-\frac{1}{\left|y-z\right|}\right|+\left|\frac{y^i-x^i}{\left|y-x\right|^2}-\frac{y^i-z^i}{\left|y-z\right|^2}\right|+\left|\frac{y^j-x^j}{\left|y-x\right|^2}-\frac{y^j-z^j}{\left|y-z\right|^2}\right| \label{first inequality}\\
& \leq & 2\left|\frac{1}{\left|y-x\right|}-\frac{1}{\left|y-z\right|}\right| + 2\left|\frac{y-x}{\left|y-x\right|^2}-\frac{y-z}{\left|y-z\right|^2}\right| \label{intermediate inequality}\\
& \leq & C\left|x-z\right|\left(\frac{1}{\left|y-u_1\right|^2}+\frac{1}{\left|y-u_2\right|^2}+\frac{1}{\left|y-u_3\right|^2}\right) \label{second inequality}.
\end{eqnarray}
By a direct computation, we have
\begin{eqnarray*}
 \mathcal{K}_{ij}(x,y)-\mathcal{K}_{ij}(z,y) &=& (\delta_{ij}-\omega_x^i \omega_z^j) \left[\frac{1}{|y-x|} - \frac{1}{|y-z|} \right]  +  \omega_z^j \left[ \frac{\omega_x^i}{|y-x|} - \frac{\omega_z^i}{|y-z|}\right]\\
 & & + \omega_x^i \left[\frac{\omega_x^j}{|y-x|} - \frac{\omega_z^j}{|y-z|} \right].
\end{eqnarray*}
Since $|\omega_x^i| \leq |\omega_x| = 1$ for all $i=1, 2, 3$ and $x \in \R^3$, we then deduce (\ref{first inequality}) and therefore (\ref{intermediate inequality}). To prove (\ref{second inequality}), we use the mean value theorem to estimate the first term of (\ref{intermediate inequality}) as 
\begin{equation}\label{first term}
\left|\frac{1}{\left|y-x\right|}-\frac{1}{\left|y-z\right|}\right| = \frac{1}{|y-u_1|^2} |(x-z) \cdot \omega_{u_1}| \leq  \frac{|x-z|}{|y-u_1|^2},
\end{equation}
for some $u_1$ in the line segment between $x$ and $z$. As for the remaining term of (\ref{intermediate inequality}), we proceed as follows. If $y$ lies in the line segment between $x$ and $z$, then 
\[|y-x| \leq |x-z|, \quad |y-z|\leq |x-z|, \quad \mbox{and}\quad |y-u_2| < |y-x|, \quad |y-u_3| < |y-z|\]
for some $u_2$ between $x$ and $y$, and some $u_3$ between $y$ and $z$, so that it is straightforward to check
\[ \left| \frac{y-x}{|y-x|^2} -\frac{y-z}{|y-z|^2} \right|\leq |x-z| \left[\frac{1}{|y-u_2|^2} + \frac{1}{|y-u_3|^2}  \right]. \]
On the other hand, if $y$ lies outside the line segment between $x$ and $z$, then 
\begin{eqnarray*}
\frac{y-x}{\left|y-x\right|^2}-\frac{y-z}{\left|y-z\right|^2} &=& \int_0^1 \frac{d}{dt} \left[ \frac{y-u(t)}{|y-u(t)|^2}\right] dt \\
&=& \int_0^1 \left[\frac{\left(2\dot{u}(t)\cdot(y-u(t))\right)(y-u(t))}{|y-u(t)|^4} -\frac{\dot{u}(t)}{|y-u(t)|^2} \right] dt,
\end{eqnarray*}
where $u(t)$ is the vector-valued function defined by 
\[ u(t)=tx+(1-t)z, \quad t\in [0,1].\]
Therefore, $\dot{u}(t)=x-z$ and we have that
\[ \left| \frac{y-x}{|y-x|^2} -\frac{y-z}{|y-z|^2} \right|\leq 3|x-z| \int_0^1 \frac{dt}{|y-u(t)|^2}.\]
Note that $u(t)$ lies in the line segment between $x$ and $z$, for all values of $t\in [0,1]$.  By continuity, there exists some $\bar{t}\in[0,1]$ such that $\max_{t\in [0,1]} |y-u(t)|^{-2} \leq |y-u(\bar{t})|^{-2} = |y-u_2|^{-2}$, $u_2:=u(\bar{t})$. This implies that 
\begin{equation}
\label{ref later on}
\left| \frac{y-x}{\left|y-x\right|^2}-\frac{y-z}{\left|y-z\right|^2}\right| \leq 3\frac{|x-z|}{|y-u_2|^2},
\end{equation}
and then conclude the proof of (\ref{second inequality}).

\smallskip

Now, since $\left|v_A\right|\leq 1$, Lemma \ref{Pallard} implies
\begin{eqnarray}
\label{Continuous Map}
 \left|T[A](x)-T[A](z)\right| & \leq & \frac{1}{2}\int_{\mathbb{R}^3}\left|\mathcal{K}(x,y)-\mathcal{K}(z,y)\right|\rho(y)dy\nonumber\\
 & \leq & C\left|x-z\right|\left\|\int_{\mathbb{R}^3}\rho(y)\frac{dy}{\left|y-\cdot\right|^2}\right\|_{L^{\infty}_x}\nonumber\\
 & \leq & C(\rho)\left|x-z\right|.
\end{eqnarray}  
Thus, $T[A]$ is a continuous vector valued function. Also, it is a simple consequence of Lemma \ref{Pallard} that
\begin{equation}
\label{Bounded Map}
 \left\|T[A]\right\|_{L^{\infty}_x}\leq 3(4\pi)^{1/3}\left\|\rho\right\|^{2/3}_{L^1_x}\left\|\rho\right\|^{1/3}_{L^{\infty}_x}\equiv \bar{C}.
\end{equation}
Therefore, $T[A]\in\mathcal{D}_{\bar{C}}$ as claimed.

Now, by virtue of the Schauder fixed point theorem \cite[Theorem 3, Section 9.1]{{McOwen}}, $T$ has a fixed point $A_{\infty}\in \mathcal{D}_{\bar{C}}$ if $T$ is a continuous mapping and the closure of the image of $T$ is compact in $\mathcal{D}_{\bar{C}}$. To show continuity, suppose that $A_k\rightarrow A$ in $\mathcal{D}_{\bar{C}}$. Since the mapping $g\mapsto v(g)=g\left(1+\left|g\right|^2\right)^{-1/2}$ is $C^1_b$, by Lemma \ref{Pallard} we have
\begin{eqnarray}
\left|T[A_k](x)-T[A](x)\right| & \leq & C\int_{\mathbb{R}^3}\int_{\mathbb{R}^3}\left|v_{A_k}-v_A\right|f(y,p)\frac{dpdy}{\left|y-x\right|}\nonumber\\
& \leq & C(\rho)\left\|A_k-A\right\|_{L^{\infty}_x}.\nonumber \end{eqnarray}
To show that $\overline{T\mathcal{D}_{\bar{C}}}\subset \mathcal{D}_{\bar{C}}$ is compact, we first notice that for $A\in \mathcal{D}_{\bar{C}}$,
\begin{equation} 
\label{Decay}
\left|T\left[A\right](x)\right|\leq \left\|\rho\right\|_{L^{\infty}_x}\int_{\texttt{supp}\,\rho}\frac{dy}{\left|x-y\right|}\leq C(\rho)\frac{1}{1+\left|x\right|}.
\end{equation}
Consider the sequence $\left\{B_{n}\right\}\subset T\mathcal{D}_{\bar{C}}$ and let $R>0$ be fixed. By (\ref{Continuous Map}) and (\ref{Bounded Map}), the restriction $$\left.\left\{B_{n}\right\}\right|_{\left\{x\in\mathbb{R}^3:\left|x\right|\leq R\right\}}$$ is equicontinuous and bounded. Then, by Arzel{\`a}-Ascoli's theorem and a standard diagonal argument we can find a subsequence $\left\{B_{n_k}\right\}$ and a continuous, bounded limit vector field $B$ such that $\left\{B_{n_k}\right\}\rightarrow B$ uniformly on compact sets, and in particular pointwise. Clearly, $\left\|B\right\|_{L^{\infty}_x}\leq \bar{C}$, and since $\left\{B_{n_k}\right\}$ satisfies the estimate (\ref{Decay}), so does $B$. We only need to show that the convergence $\left\{B_{n_k}\right\}\rightarrow B$ is uniform. Indeed, let $\epsilon>0$. Choose $R>0$ such that the right-hand side of (\ref{Decay}) is less than $\epsilon/2$ for $\left|x\right|>R$. Then, for all $k$ we have $\left|B_{n_k}(x)-B(x)\right|<\epsilon$ for $\left|x\right|>R$, and we can find a $k_0=k_0(R,\epsilon)$ such that for all $k>k_0$
$$ \sup_{\left|x\right|\leq R}\left|B_{n_k}(x)-B(x)\right|<\epsilon.
$$
This proves uniform convergence. Hence, all the hypotheses for the Schauder fixed point theorem are fulfilled, and thus $T$ has a fixed point $A_{\infty}$ in $\mathcal{D}_{\bar{C}}$. This completes the existence proof.

Finally, we show that $A_{\infty}$ has the required regularity. To that end, define $v_{A_{\infty}}$ and then $j_{A_{\infty}}$ according to (\ref{Vlasov In RVD}) and (\ref{Densities}), respectively. Since $\left|v_{A_{\infty}}\right|\leq1$, then $\left|j_{A_{\infty}}\right|\leq\rho$. Thus, $j_{A_{\infty}}\in L^\infty (\mathbb{R}^3;\mathbb{R}^3)$ has compact support. On the other hand, $\left|\mathcal{K}(x,y)\right|\leq 2\left|y-x\right|^{-1}$, and since the $imk$-th entry of $\partial_x\mathcal{K}$ reads
\begin{eqnarray*}
\left(\partial_x\mathcal{K}\right)_{imk}(x,y) & = & \partial_{x_k}\left\{\left|y-x\right|^{-1}\left[\delta_{im}+\omega^i\omega^m\right]\right\}\\
& = & \left|y-x\right|^{-2}\left[\delta_{im}\omega^k-\delta_{km}\omega^i-\delta_{ik}\omega^m+3\omega^i\omega^k\omega^m\right],
\end{eqnarray*}
the kernel $\mathcal{K}(x,y)$ satisfies the derivative estimate $\left|\partial_x\mathcal{K}(x,y)\right|\leq 6\left|y-x\right|^{-2}$. Hence, we can use the standard theory for Poisson's equation to find that $\partial_xA_{\infty}\in C(\mathbb{R}^3;\mathbb{R}^3)$ as claimed; see, for instance, \cite[Lemma 4.1]{Trudinger} or \cite[Theorem 10.2 (iii)]{Lieb}. The remaining assertions in Lemma \ref{Lemma Properties Potentials} are easy to check. In particular, since $\left|v_{A_{\infty}}\right|\leq1$ and $\rho\in L^{\infty}(\mathbb{R}^3;\mathbb{R})$, we have by Lemma \ref{Pallard} that $\partial_xA_{\infty}$ is bounded. This completes the proof of the lemma.
\qed
 
 \smallskip
 \smallskip
 
 The potentials satisfy the following estimates:   
\begin{lemma} 
\label{Estimate Darwin Potentials}
Let $f$ be given as in Definition \ref{Definition Solution} and define $\rho$ according to (\ref{Densities}). Suppose that $\Phi$ and $A$ are as in Lemma \ref{Lemma Properties Potentials}. If $\rho\in L^{\infty}(\mathbb{R}^3;\mathbb{R})$, then 
\begin{equation}
\label{LINFTY Derivative Potential}
\left\|A\right\|_{L^{\infty}_x}\leq C\left\|\rho\right\|^{2/3}_{L^1_x}\left\|\rho\right\|^{1/3}_{L^{\infty}_x},\quad
\left\|\partial_xA\right\|_{L^{\infty}_x}\leq C\left\|\rho\right\|^{1/3}_{L^1_x}\left\|\rho\right\|^{2/3}_{L^{\infty}_x}.
\end{equation}
These estimates also hold for the scalar potential $\Phi$. Moreover, there exists a positive constant $C$ that depends on $\left\|\rho\right\|_{L^1_x}$ and $\left\|\rho\right\|_{L^{\infty}_x}$ such that ,
\begin{equation}
\label{Lipschitz}
\left|A(x)-A(z)\right|+\left|\partial_xA(x)-\partial_xA(z)\right|+\left|\partial_x\Phi(x)-\partial_x\Phi(z)\right|\leq -C\left|x-z\right|\ln\left|x-z\right|.
\end{equation}
for any $\left(x,z\right)\in\mathbb{R}^3\times\mathbb{R}^3$ with $\left|x-z\right|\leq1/2$.
 \end{lemma}
 
{\em Proof.} These are standard results for the scalar potential $\Phi$ which were already used in \cite{Loeper} to prove the uniqueness of solutions to the Vlasov-Poisson system. Therefore, we only work here with the vector potential $A$. Following the notation in the proof of Lemma \ref{Lemma Properties Potentials}, we have that $\left|\mathcal{K}(x,y)\right|\leq 2\left|y-x\right|^{-1}$ and $\left|\partial_x\mathcal{K}(x,y)\right|\leq 6\left|y-x\right|^{-2}$. Then, since $\left|v_{A}\right|\leq1$ and thus $\left|j_{A}\right|\leq\rho$, the estimates in (\ref{LINFTY Derivative Potential}) readily follow by Lemma \ref{Pallard}.

To prove (\ref{Lipschitz}) we rely on a similar result discussed in \cite[Lemma 8.1]{Majda} for the $2$D Euler equation. To begin with, let $h=\left|x-z\right|\leq1/2$ and $B_{r}(x)$ be a ball of radius $r$ centered at $x$. A computation as for (\ref{first inequality})-(\ref{second inequality}) in the proof of Lemma \ref{Lemma Properties Potentials} shows that the kernel $\mathcal{K}(x,y)$ and its derivative satisfy
$$ \left|\mathcal{K}(x,y)-\mathcal{K}(z,y)\right|\leq C\left(\left|\frac{y-x}{\left|y-x\right|^2}-\frac{y-z}{\left|y-z\right|^2}\right|+\left|\frac{1}{\left|y-x\right|}-\frac{1}{\left|y-z\right|}\right|\right)
$$
and 
$$ \left|\partial_y\mathcal{K}(x,y)-\partial_y\mathcal{K}(z,y)\right|\leq C\left(\left|\frac{y-x}{\left|y-x\right|^3}-\frac{y-z}{\left|y-z\right|^3}\right|+\left|\frac{1}{\left|y-x\right|^2}-\frac{1}{\left|y-z\right|^2}\right|\right).
$$
As a result,
\begin{eqnarray*}
\lefteqn{ \left|A(x)-A(z)\right|+\left|\partial_xA(x)-\partial_xA(z)\right|}\\
& \leq & \int_{\mathbb{R}^3}\left(\left|\mathcal{K}(x,y)-\mathcal{K}(z,y)\right| +\left|\partial_y\mathcal{K}(x,y)-\partial_y\mathcal{K}(z,y)\right| \right) \rho(y)dy\\
& \leq & \int_{\mathbb{R}^3}\left(\left|\frac{y-x}{\left|y-x\right|^3}-\frac{y-z}{\left|y-z\right|^3}\right|+
         \left|\frac{y-x}{\left|y-x\right|^2}-\frac{y-z}{\left|y-z\right|^2}\right|\right.\\
&      &  \left.+ \left|\frac{1}{\left|y-x\right|^2}-\frac{1}{\left|y-z\right|^2}\right| + \left|\frac{1}{\left|y-x\right|}-\frac{1}{\left|y-z\right|}\right|\right)\rho(y)dy\\
& = &  I+II+III+IV.
\end{eqnarray*}
In the remainder of the proof we shall only estimate $I$ since it is slightly more involved than the other three integrals and they can all be estimated in the same fashion. To proceed, consider
\begin{eqnarray*}
I & = & \left[\int_{B_{2h}(x)}+\int_{B_2(x)/B_{2h}(x)}+\int_{\mathbb{R}^3/B_2(x)}\right]\left|\frac{y-x}{\left|y-x\right|^3}-\frac{y-z}{\left|y-z\right|^3}\right|\rho(y)dy\\
& = & I_1+I_2+I_3.
\end{eqnarray*}
We estimate one integral at a time. 
\begin{eqnarray*}
I_1 & \leq &  \left\|\rho\right\|_{L^{\infty}_x}\left(\int_{B_{2h}(x)}\frac{dy}{\left|y-x\right|^2}+\int_{B_{2h}(x)}\frac{dy}{\left|y-z\right|^2}\right)\\
& \leq & C\left\|\rho\right\|_{L^{\infty}_x}\left(\int^{2h}_0dr+\int^{3h}_0dr\right)\: \leq\:  C\left\|\rho\right\|_{L^{\infty}_x}\left|x-z\right|.
\end{eqnarray*}
As for $I_2$, let $y\in B_2(x)/B_{2h}(x)$. Similar to the estimate (\ref{ref later on}) in the proof of Lemma \ref{Lemma Properties Potentials}, we can show that  
$$ \left|\frac{y-x}{\left|y-x\right|^3}-\frac{y-z}{\left|y-z\right|^3}\right|\leq C\frac{\left|x-z\right|}{\left|y-u\right|^3} 
$$
for some $u$ on the line segment between $x$ and $z$. Then, since for some constant $C>0$ we have $\left|y-x\right|\leq C\left|y-u\right|$, 
\begin{eqnarray*}
I_2 & \leq & C\left\|\rho\right\|_{L^{\infty}_x}\left|x-z\right|\int_{B_{2}(x)/B_{2h}(x)}\frac{dy}{\left|y-x\right|^3}\\
& \leq & C\left\|\rho\right\|_{L^{\infty}_x}\left|x-z\right|\int^{2}_{2h}\frac{dr}{r}\: \leq\: -C\left\|\rho\right\|_{L^{\infty}_x}\left|x-z\right|\ln\left|x-z\right|.
\end{eqnarray*}
To estimate $I_3$, let $y\in \mathbb{R}^3/B_2(x)$. Then $\left|y-x\right|\geq2$ and we use the mean value theorem and a standard estimate \cite[Lemma 8.1]{Majda} to find that 
\begin{eqnarray*}
\lefteqn{\left|\frac{y-x}{\left|y-x\right|^3}-\frac{y-z}{\left|y-z\right|^3}\right|}\\
& \leq & \frac{1}{\left|y-x\right|}\left|\frac{1}{\left|y-x\right|}-\frac{1}{\left|y-z\right|}\right|+\frac{1}{\left|y-z\right|} \left|\frac{y-x}{\left|y-x\right|^2}-\frac{y-z}{\left|y-z\right|^2}\right|\\
& \leq & \frac{\left|x-z\right|}{\left|y-x\right|\left|y-u\right|^2}+\frac{\left|x-z\right|}{\left|y-x\right|\left|y-z\right|^2}\:\leq\: \frac{1}{2}\left|x-z\right|\left(\frac{1}{\left|y-u\right|^2}+\frac{1}{\left|y-z\right|^2}\right)
\end{eqnarray*}
for some other $u$ on the line segment between $x$ and $z$. Hence, we have by Lemma \ref{Pallard},
$$ I_3\leq \left|x-z\right|\left\|\int_{\mathbb{R}^3}\rho(y)\frac{dy}{\left|y-\cdot\right|^2}\right\|_{L^{\infty}_x}\leq C\left\|\rho\right\|^{\frac{1}{3}}_{L^1_x}\left\|\rho\right\|^{\frac{2}{3}}_{L^{\infty}_x}\left|x-z\right|.
$$
We gather these estimates and use the fact that $|x-z|\leq 1/2$ to find that for some constant $C(\rho)$ that depends on $\left\|\rho\right\|_{L^1_x}$ and $\left\|\rho\right\|_{L^{\infty}_x}$, 
\begin{equation}
\label{Estimate}
I\leq-C(\rho)\left|x-z\right|\ln\left|x-z\right|.
\end{equation}
Thus, since the same rationale shows that (\ref{Estimate}) also holds for the integrals $II$, $III$ and $IV$, we conclude that 
$$
\left|\partial_xA(x)-\partial_xA(z)\right|\leq -C(\rho)\left|x-z\right|\ln\left|x-z\right|,
$$
and the proof of the lemma is complete. 
\qed
 
\smallskip
 
We conclude this section with a lemma that characterizes the weak solutions of the RVD system via the associated characteristic system. We recall that the speed of light has been set to $c=1$.

\begin{lemma}
\label{Definition Linear Vlasov}
Let $\Phi$ and $A$ be given as in Lemma \ref{Lemma Properties Potentials}. Then, there exists a unique solution $(X,P)(s,t,x,p)$, $0\leq s\leq t<T$, to the characteristic system 
\begin{eqnarray}
\label{Characteristics X}
\dot{x} & = & v_A(s,x,p),\\
\label{Characteristics P}
\dot{p} & = & \left[-\nabla\Phi+v^i_A\nabla A^i\right](s,x,p)
\end{eqnarray} 
with $(X,P)(t,t,x,p)=(x,p)$, associated to the (linear!) Vlasov equation (\ref{Vlasov In RVD}). Moreover, since the right-hand side of (\ref{Characteristics X})-(\ref{Characteristics P}) is an incompressible vector field, the mapping $(x,p) \mapsto (X,P)$ is measure preserving. Conversely, the weak solution of (\ref{Vlasov In RVD}) in the sense of (\ref{Solution Distributions}) with $f_0\in L^1\cap L^{\infty}(\R^6;\mathbb{R})$, $f_0\geq0$, is uniquely determined by $f(t,x,p)=f_0((X,P)(0,t,x,p))$ on $[0,T[$ for all $(x,y)\in B\subset\mathbb{R}^6$, $B$ a Borel set.   
\end{lemma}

{\em Proof}. 
Set $z=(x,p)$. Since the mapping $g\mapsto v(g)=g\left(1+\left|g\right|^2\right)^{-1/2}$ is $C^1_b$, Lemma \ref{Estimate Darwin Potentials} guarantees that the vector field $G=\left(v_A,-\nabla\Phi+v^i_A\nabla A^i\right)$ on the right-hand side of the equations (\ref{Characteristics X})-(\ref{Characteristics P}) is Lipschitz continuous in the momentum variable and Log-Lipschitz in space. This implies that there exists a unique solution $[0,T[\ni s\mapsto Z=(X,P)(s,\cdot,\cdot,\cdot)$ to the characteristic system (\ref{Characteristics X})-(\ref{Characteristics P}) and the characteristic flow $Z(s,t,z)$ is H{\"o}lder continuous with respect to $z$ \cite[Chapter 8]{Majda}. Moreover, since by \cite[Lemma 1]{ARMA} we have in the classical sense
\begin{equation}
\label{Divergence Free}
\nabla_z\cdot G=\nabla_x\cdot v_A+\nabla_p\cdot\left(-\nabla\Phi+v^i_A\nabla A^i\right)\equiv 0,
\end{equation}
the mapping $\mathbb{R}^6 \ni z\mapsto Z(\cdot,\cdot,z)$ --with inverse $Z^{-1}(s,t,\cdot)=Z(t,s,\cdot)$-- is measure preserving \cite[Chapter 8]{Majda}. By (\ref{Divergence Free}), we can write the Vlasov equation in divergence form, i.e., 
\begin{equation}
\label{Vlasov Divergence Free}
\partial_tf(t,z)+\nabla_{z}\cdot\left(G(t,z)f(t,z)\right)=0.
\end{equation}
Then, for $f_0$ given as in the lemma, we have (as a corollary of \cite[Theorem 8.2.1]{AGS}) that the function $f(t,z)=f_0(Z(0,t,z))$, $t\in[0,T[, \,z=(x,y)\in B\subset\mathbb{R}^6$, where $B$ is a Borel set, with $\left.f\right|_{t=0}=f_0$, is the unique solution to the equation (\ref{Vlasov Divergence Free}) (and so (\ref{Vlasov In RVD})) in the sense of (\ref{Solution Distributions}).
\qed

\smallskip

\begin{remark} 
\label{PushForwardRemark}
Since by Lemma \ref{Definition Linear Vlasov}, the weak solution $f$ of (\ref{Vlasov In RVD}) in the sense of (\ref{Solution Distributions}) satisfies $f\left(t,Z(t,0,z)\right)=f_0(z)$ where $z\mapsto Z(t,0,z)$ is measure-preserving, then for all $\varphi\in C_0(\mathbb{R}^6;\mathbb{R})$, we have
$$ \int_{\mathbb{R}^6}\varphi(z)f(t,z)dz=\int_{\mathbb{R}^6}\varphi(Z(t,0,z))f_0(z)dz,$$
which means that the map $z\mapsto Z(t,0,z)$ transports $f_0$ to $f(t)$, or $f(t,z)=Z(t,0,z)_{\#}f_0$ as defined in Section \ref{transport section}; see (\ref{pushforward}).
\end{remark}


\section{Tools from Optimal Transportation}
\label{transport section}

Denote by ${\bf P}_2(\R^3\times\R^3)$ the set of probability densities $f(x,p)$ on $\R^3\times\R^3$ with finite second moment, $\int_{\R^3\times\R^3} \left(|x|^2+|p|^2\right)f(x,p)\,\mbox{d}x\,\mbox{d}p <\infty$. The $L^2$-Wasserstein distance between two densities $f_1(x,p)$ and $f_2(x,p)$ in ${\bf P}_2(\R^3\times\R^3)$ is defined by
\begin{eqnarray}\label{def wasserstein}
\lefteqn{W_2^2(f_1,f_2)} \nonumber\\
&=& \inf\Big\{ \int_{\R^3\times\R^3}\int_{\R^3\times\R^3}|(x,p)-(y,q)|^2\,\mbox{d}\gamma\left((x,p), (y,q)\right); \;\; \gamma\in\Gamma(\mu_1, \mu_2) \Big\} \nonumber \\
& = & \inf\Big\{ \int_{\R^3}\int_{\R^3} |T(x,p)-(x,p)|^2f_1(x,p)\,\mbox{d}x\,\mbox{d}p; \quad T_{\#}f_1=f_2\Big\} 
\end{eqnarray}
where $\mbox{d}\mu_1=f_1(x,p)\mbox{d}x\,\mbox{d}p$, $\mbox{d}\mu_2=f_2(x,p)\mbox{d}x\,\mbox{d}p$, $\Gamma(\mu_1, \mu_2)$ denotes the set of all probability measures on $\R^6\times\R^6$ with marginals $\mu_1$ and $\mu_2$, and $T_{\#}f_1=f_2$ means that 
\begin{equation}\label{pushforward}
\int_{\R^3}\int_{\R^3} \varphi(x,p)f_2(x,p)\,\mbox{d}x\,\mbox{d}p =\int_{\R^3}\int_{\R^3} \varphi\left(T(x,p)\right)f_1(x,p)\,\mbox{d}x\,\mbox{d}p
\end{equation}
for all test functions $\varphi\in C_0(\R^3\times\R^3)$. In \cite{B}, Brenier proved that the minimization problem (\ref{def wasserstein}), the so-called Monge-Kantorovich problem, has a unique solution T, which is characterized $\mu_1$-a.e. by the gradient of a convex function $\phi: \R^3\times\R^3 \rightarrow \R$, i.e.,  $T$ is uniquely determined $\mu_1$-a.e. by  $T=\nabla\phi$ with $(\nabla\phi)_{\#}f_1=f_2$ for some convex function $\phi$. Note that in (\ref{def wasserstein}), the minimizers $\gamma$ and $T$ are related by $\gamma=(\texttt{id}_{\mathbb{R}^6},T)_{\#}f_1$. Moreover, if $\theta\in [1,2]$ and 
\begin{equation}\label{mccann interpolation}
f_\theta={T_{\theta}}_{\#}f_1, \quad T_\theta=(2-\theta)\texttt{id}_{\mathbb{R}^6} + (\theta-1) T = \nabla\left((2-\theta)\frac{|\cdot|^2}{2}+(\theta-1)\phi\right)
\end{equation}
denotes McCann's interpolation \cite{Mc}, then the curve $[1,2]\ni \theta \mapsto f_\theta \in {\bf P}_2(\R^3\times\R^3)$ is the unique length minimizing geodesic joining $f_1$ to $f_2$ in the Wasserstein space $\left({\bf P}_2(\R^3\times\R^3), W_2\right)$, in the sense that  $W_2(f_1,f_2)=W_2(f_1,f_\theta) + W_2(f_\theta,f_2)$. Furthermore, the interpolant $f_\theta$ satisfies the continuity equation in a weak sense, 
\begin{equation}\label{transport equation}
\partial_\theta f_\theta(x,p) +\nabla_{x,p}\cdot \left(u_\theta(x,p)f_\theta(x,p)\right)=0 \quad \forall\, \theta\in [1,2],
\end{equation}
where $u_\theta\in L^2_{f_\theta}(\R^3\times\R^3;\R^3)$ is the velocity field associated with the trajectory $f_\theta$, i.e.,
\begin{equation}\label{velocity field}
u_\theta\left(T_\theta(x,p)\right) = \frac{\partial T_\theta(x,p)}{\partial\theta} = \nabla\phi(x,p)-(x,p).
\end{equation}
Indeed, (\ref{transport equation}) can be formally seen as follows. For any test function $\varphi\in C^1_0(\R^3\times\R^3)$, using (\ref{pushforward}) with $f_\theta=(T_\theta)_{\#}f_1$ and then (\ref{velocity field}), we have:
\begin{eqnarray}
\lefteqn{\frac{d}{d\theta} \int_{\R^3}\int_{\R^3} \varphi(x,p) f_\theta(x,p)\,\mbox{d}x\,\mbox{d}p}\nonumber\\
 &=& \int_{\R^3}\int_{\R^3} \nabla\varphi\left(T_\theta(x,p)\right)u_\theta\left(T_\theta(x,p)\right)f_1(x,p)\,\mbox{d}x\,\mbox{d}p \nonumber\\
&=& \int_{\R^3}\int_{\R^3} \nabla\varphi(x,p)u_\theta(x,p) f_\theta(x,p)\,\mbox{d}x\,\mbox{d}p \label{weak transport}\\
&=& -\int_{\R^3}\int_{\R^3} \varphi(x,p)\nabla_{x,p}\cdot\left(f_\theta(x,p) u_\theta(x,p)\right)\,\mbox{d}x\,\mbox{d}p, \nonumber
\end{eqnarray}
where we use an integration by parts on the right-hand side integral in (\ref{weak transport}).  

Combining (\ref{def wasserstein}) - (\ref{velocity field}), we have that
\begin{eqnarray}\label{another wasserstein}
W_2^2(f_1,f_2)
 &=& \int_{\R^3}\int_{\R^3} |\nabla\Phi(x,p)-(x,p)|^2f_1(x,p)\,\mbox{d}x\,\mbox{d}p\nonumber\\
& = & \int_{\R^3}\int_{\R^3} |u_\theta\left(T_\theta(x,p)\right)|^2f_1(x,p)\,\mbox{d}x\,\mbox{d}p \nonumber\\ 
 &=& \int_{\R^3}\int_{\R^3} |u_\theta(x,p)|^2f_\theta(x,p)\,\mbox{d}x\,\mbox{d}p.
\end{eqnarray}
 Formula (\ref{another wasserstein}) is commonly known as the Benamou-Brenier \cite{BB} characterization of the $L^2$-Wassertein distance, namely,
\begin{eqnarray}\label{benamou-brenier}
\lefteqn{W_2^2(f_1,f_2)}\nonumber\\
 & = & \min\Big\{\int_1^2 \int_{\R^3}\int_{\R^3}f(\theta,x,p)|u(\theta,x,p)|^2\mbox{d}x\,\mbox{d}p\,\mbox{d}\theta; \;\; f(\theta)\in {\bf P}_2(\R^3\times\R^3)  \Big\},\nonumber
\end{eqnarray}
where the minimum is taken over all  absolutely continuous curves $f: [1,2]\ni\theta \mapsto f(\theta)\in {\bf P}_2(\R^3\times\R^3)$ satisfying the constraints $ f(1)=f_1, \; f(2)=f_2$ and $ \partial_\theta f+ \nabla_{x,p}\cdot(uf)=0$ . For a development on this topic, we refer to \cite{AGS}.

\smallskip

In the next lemma, we collect some well-known results in optimal transport theory that will be needed later in the paper. 

\begin{lemma}  \label{Lemma optimal transport}
Let $f_1, f_2\in L^\infty(\R^3\times\R^3;\mathbb{R})$ be two probability densities in ${\bf P}_2(\R^3\times\R^3)$ with compact supports. For any $\theta\in [1,2]$, define the interpolant $f_\theta$ as in (\ref{mccann interpolation}). Then
\begin{enumerate}
\item For all $\theta\in [1,2]$, $f_\theta$ has a compact support in $\R^3\times\R^3$, and 
\begin{equation}\label{f theta estimate}
\|f_\theta\|_{L^\infty_{x,p}} \leq \max\{ \|f_1\|_{L^\infty_{x,p}}, \|f_2\|_{L^\infty_{x,p}}\}.
\end{equation}
That is,  $[1,2]\ni \theta \mapsto f_\theta$ belongs to $L^\infty \left([1,2], L^1\cap L^\infty(\R^3\times\R^3);\R\right)$.
\item Moreover, $[1,2]\ni \theta \mapsto f_\theta$ is differentiable at every point $\theta\in [1,2]$, and its derivative $\partial_\theta f_\theta$, defined in the weak sense by (\ref{weak transport}), satisfies
\begin{equation}\label{derivative f theta estimate}
\|\partial_\theta f_\theta \|_{H^{-1}(\R^3\times\R^3)} \leq \max\{ \|f_1\|_{L^\infty_{x,p}}, \|f_2\|_{L^\infty_{x,p}}\}^{1/2} \, W_2(f_1, f_2), \; \forall\, \theta\in [1,2].
\end{equation}
That is, $[1,2]\ni \theta \mapsto \partial_\theta f_\theta$ belongs to $L^\infty\left([1,2], H^{-1}(\R^3\times\R^3);\R\right)$.
\end{enumerate} 
\end{lemma}

{\em Proof.} The proofs of (\ref{f theta estimate}) and (\ref{derivative f theta estimate}) are done in \cite{Loeper}. Here we only show that $f_\theta$ has a compact support in $\R^3\times\R^3$. Indeed, assume that the support of $f_i$, $\texttt{supp}f_i$, is contained in the ball $B_{R_i}$ centered at the origin with radius $R_i$ for $i=1, 2$. By the definition (\ref{mccann interpolation}) of $f_\theta$, $T_\theta$ is the optimal map in $W^2_2(f_1,f_\theta)$. Then for $\mu_1$-a.e., $T_\theta$ is invertible and $\partial_{x,p} T_\theta(x,p)$ is diagonalizable  with positive eigenvalues (see \cite[Thm 6.2.4 \& Prop 6.2.12]{AGS}). Moreover, the following Monge-Amp\`ere equation holds for $\mu_1$-a.e. $(x,p)\in \R^3\times\R^3$,
\[f_1(x,p)=f_\theta\left(T_\theta(x,p)\right)\,\texttt{det}\partial_{x,p}T_\theta(x,p).\]
It follows that for $\mu_1$-a.e. $(x,p)\in \R^3\times\R^3$, if $f_\theta\left(T_\theta(x,p)\right) \neq 0$, then $f_1(x,p)\neq 0$ so that 
\[\{(x,p):\; f_\theta(x,p)\neq 0\} \subset T_\theta \left(\{ (x,p):\; f_1(x,p)\neq 0\}\right)\]
i.e. $\texttt{supp}f_\theta \subset \overline{T_\theta(B_{R_1})}$. But $T_\theta(B_{R_1}) \subset (\theta-2)B_{R_1} + (\theta-1)B_{R_2} \subset  B_{R}$ where $R = R_1 + R_2$. Therefore $\texttt{supp}f_\theta \subset \overline{B_R}$, i.e. $f_\theta$ has a compact support.
\qed


\section{Final Estimates}
\label{Estimates on the Field}

We first note that if $f_0\in L^1\cap L^\infty(\R^3\times\R^3;\R)$ is such that $f_0\geq 0$, and if $f$ is a weak solution of the RVD system in the sense of Definition \ref{Definition Solution}, then  $f(t) \geq 0$ and $\|f(t)\|_{L^1_{x,p}} = \|f_0\|_{L^1_{x,p}}$ for all $t\in [0,T[$, so that $f(t)$ can be viewed as a probability density on $\R^3\times\R^3$ up to normalizing the $L^1$-norm of $f_0$ to $1$. Moreover, under the assumption that the support of $f(t)$ is compact in $\R^3\times\R^3$ for all $t\in [0,T[$, then $f(t)\in {\bf P}_2(\R^3\times\R^3)$.

\smallskip

For simplicity and without loss of generality, we shall omit the time dependence in $f$ throughout this section. The next lemma gives estimates on the scalar and vector potentials induced by two bounded probability densities.

\begin{lemma} 
\label{Potentials Estimate Lemma 1}
Let $f_1, f_2 \in {\bf P}_2\cap L^\infty(\R^3\times\R^3;\R)$. Define $\rho_i$ and $j_{A_i}$ according to (\ref{Densities}), and let $\Phi_i$ and $A_i$, $i=1,2$, satisfy respectively the equations
\begin{equation}
\label{Equation Scalar Potential Lemma} 
\Delta\Phi_i(x) = -4\pi\rho_i(x),\quad  \lim_{\left|x\right|\rightarrow\infty}\Phi_i(x) = 0
\end{equation}
and
\begin{equation}
\label{Equation Vector Potential Lemma} 
\Delta A_i(x) =  -4\pi j_{A_i}(x)-\nabla\left\{\nabla\cdot\int_{\mathbb{R}^3}j_{A_i}(y)\frac{dy}{\left|y-x\right|}\right\}, \quad \lim_{\left|x\right|\rightarrow\infty} \left|A_i(x)\right| = 0,
\end{equation}
in the sense of (\ref{Potentials Phi})-(\ref{Densities}) (see Lemma \ref{Lemma Vector Potential}). Assume that for some $R>0$, $\texttt{supp}f_1\cup\texttt{supp}f_2\subset B_R\times B_R$. Then there exists a constant $C>0$ which depends on $R$, $\left\|f_1\right\|_{L^{\infty}_{x,p}}$ and $\left\|f_2\right\|_{L^{\infty}_{x,p}}$, such that
\begin{equation}
\label{Potentials Estimate} \left\|\partial_x\Phi_1-\partial_x\Phi_2\right\|_{L^2_x}+\left\|\partial_xA_1-\partial_xA_2\right\|_{L^2_x}\leq C \,W_2(f_1,f_2),
\end{equation}
and, for $i\in \{1,2\}$,
\begin{equation}
\label{Last Integral}
 \int_{\mathbb{R}^3}\rho_i(x)\left|A_1(x)-A_2(x)\right|^2dx\leq C \,W^2_2(f_1,f_2).
\end{equation}
\end{lemma}

{\em Proof}.  The estimate on the scalar potential, that is, the first term on the left-hand side of (\ref{Potentials Estimate}), is essentially proved in \cite{Loeper} under the weaker assumption of the boundedness of the charge density. Here we only prove the estimates on the vector potential, which are (\ref{Last Integral}) and the second term on the left-hand side of (\ref{Potentials Estimate}). The first part of the proof follows closely the steps in the uniqueness result of Lemma \ref{Lemma Properties Potentials}, except that here we are given two densities $f_1$ and $f_2$. 

Indeed, for $i \in \{1, 2\}$, set $g_{A_i}(x,p)= p-A_i(x)$ and 
 \[j_{A_i}(x)=\int_{\R^3} v_{A_i}(x,p)f_i(x,p)\,\mbox{d}p, \quad  v_{A_i}=\frac{g_{A_i}}{\sqrt{1+\left|g_{A_i}\right|^2}}.\]
By the assumptions on $f_i$, we have that $j_{A_i}\in L^1\cap L^\infty(\R^3;\R^3)$ and its support is included in $B_R$.  Then, $j_{A_i}\in L^2(\R^3;\R^3)$, and we can argue as in the uniqueness proof of Lemma \ref{Lemma Properties Potentials} to conclude that $A_i\in W^{1,2}(\R^3;\R^3)$. Taking the difference of the two equations in (\ref{Equation Vector Potential Lemma}), we obtain
\[-\Delta (A_1-A_2)(x) = 4\pi \left(j_{A_1}-j_{A_2}\right)(x) + \nabla\left\{\nabla\cdot\int_{\mathbb{R}^3} \left(j_{A_1}-j_{A_2}\right)(y)\frac{dy}{\left|y-x\right|}\right\},\]
and since $A_1-A_2\in W^{1,2}(\R^3;\R^3)$, integration by parts against $A_1-A_2$ yields
\begin{eqnarray*}
\int_{\R^3} \left|\partial_xA_1 - \partial_xA_2\right|^2\,\mbox{d}x  &=& 4\pi \int_{\R^3} (A_1-A_2)\cdot (j_{A_1}-j_{A_2})\,\mbox{d}x \\
& & \quad - \int_{\R^3}\nabla\cdot(A_1-A_2)\nabla\cdot I\,\mbox{d}x,
\end{eqnarray*}
where 
\[ I(x) = \int_{\mathbb{R}^3} \left(j_{A_1}-j_{A_2}\right)(y)\frac{dy}{\left|y-x\right|}.\]
By Lemma \ref{Lemma Properties Potentials}, we have $\nabla\cdot A_i=0$. Thus the last integral on the right-hand side vanishes. On the other hand, we write
 \begin{eqnarray*}
j_{A_1}-j_{A_2} &=& \int_{\R^3} f_1(v_{A_1}-v_{A_2})\,\mbox{d}p + \int_{\R^3} v_{A_2} (f_1-f_2)\,\mbox{d}p \\
&=& \int_{\R^3} f_1(v_{A_1}-v_{A_2})\,\mbox{d}p - \int_{\R^3} \int_1^2 v_{A_2}\partial_\theta f_\theta \,\mbox{d}\theta\,\mbox{d}p,
\end{eqnarray*}
where $f_\theta$ is the interpolant (\ref{mccann interpolation}) between $f_1$ and $f_2$. Inserting this identity into the above expression, we have
 \begin{eqnarray*}
\int_{\R^3} \left|\partial_xA_1 - \partial_xA_2\right|^2\,\mbox{d}x - 4\pi \int_{\R^3}\int_{\R^3} f_1(A_1-A_2)\cdot (v_{A_1}-v_{A_2})\,\mbox{d}p\,\mbox{d}x\\
\quad = \quad -4\pi \int_1^2  \int_{\R^3}\int_{\R^3} (A_1-A_2)\cdot v_{A_2} \partial_\theta f_\theta \,\mbox{d}p\,\mbox{d}x\,\mbox{d}\theta.
 \end{eqnarray*}
Therefore,
\begin{eqnarray}
\label{Estimate Three Integrals}
\lefteqn{\int_{\R^3} \left|\partial_xA_1 - \partial_xA_2\right|^2\,\mbox{d}x - 4\pi \int_{\R^3}\int_{\R^3} f_1(A_1-A_2)\cdot (v_{A_1}-v_{A_2})\,\mbox{d}p\,\mbox{d}x} \nonumber\\
& &= \quad 4\pi\int_1^2  \int_{\R^3}\int_{\R^3} f_\theta u_\theta^x\cdot \partial_x\left[(A_1-A_2)\cdot v_{A_2}\right] \,\mbox{d}p\,\mbox{d}x\,\mbox{d}\theta\nonumber \\
& & \quad + \; 4\pi\int_1^2  \int_{\R^3}\int_{\R^3} f_\theta u_\theta^p\cdot \partial_p\left[(A_1-A_2)\cdot v_{A_2}\right] \,\mbox{d}p\,\mbox{d}x\,\mbox{d}\theta,
\end{eqnarray}
where we use the continuity equation (\ref{transport equation}) with the velocity field $u_\theta$ denoted by $u_{\theta}=(u^x_{\theta},u^p_{\theta})$. Here $u_\theta^x$ and $u_\theta^p$ are the $x$ and $p$-components in $\R^3$ of $u_\theta$, respectively. We now estimate each of the integral terms in (\ref{Estimate Three Integrals}). 

Indeed, as in the uniqueness proof of Lemma \ref{Lemma Properties Potentials}, we can deduce that for some $C_R>0$, the estimate
\[(A_1-A_2)\cdot (v_{A_1}-v_{A_2}) \leq - C_R |A_1 - A_2|^2\]
holds uniformly on $B_R\times B_R$. Then, the left-hand side of (\ref{Estimate Three Integrals}) is bounded below as
\begin{eqnarray}\label{lhs bound}
\int_{\R^3} \left|\partial_xA_1 - \partial_xA_2\right|^2\,\mbox{d}x - 4\pi \int_{B_R}\int_{B_R} f_1(A_1-A_2)\cdot (v_{A_1}-v_{A_2})\,\mbox{d}p\,\mbox{d}x \nonumber\\
\quad  \geq C_R \; \left( \left\|\partial_xA_1-\partial_xA_2\right\|^2_{L^2_x} + \left\|\rho_1^{1/2}(A_1-A_2)\right\|^2_{L^2_x} \right).
\end{eqnarray}
On the other hand, by inserting the identities $\partial_x\left[(A_1-A_2)\cdot v_{A_2}\right] = v_{A_2}\cdot (\partial_xA_1-\partial_xA_2)+(A_1-A_2)\cdot \partial_x v_{A_2}$ and $\partial_p\left[(A_1-A_2)\cdot v_{A_2}\right] = (A_1-A_2)\cdot \partial_p v_{A_2}$ into (\ref{Estimate Three Integrals}), it is easy to see that the expression on the right-hand side of  (\ref{Estimate Three Integrals}) is dominated by 
\begin{eqnarray*}
I_1 + I_2 + I_3 & = & 4\pi\int_1^2\int_{\mathbb{R}^3}\int_{\mathbb{R}^3}f_{\theta}\left|u_{\theta}\right|\left|\partial_xA_1-\partial_xA_2\right|\,\mbox{d}x\,\mbox{d}p\,\mbox{d}\theta \nonumber\\
& & +\; 4\pi\int_1^2\int_{\mathbb{R}^3}\int_{\mathbb{R}^3}f_{\theta}\left|u_{\theta}\right|\left|A_1-A_2\right|\left|\partial_xA_2\right|\,\mbox{d}x\,\mbox{d}p\,\mbox{d}\theta \nonumber\\
& & + \; 4\pi\int_1^2\int_{\mathbb{R}^3}\int_{\mathbb{R}^3}f_{\theta}\left|u_{\theta}\right|\left|A_1-A_2\right|\,\mbox{d}x\,\mbox{d}p\,\mbox{d}\theta .
\end{eqnarray*}
Since by Lemma \ref{Lemma optimal transport} we have $\|\rho_\theta\|_{L^\infty_x} \leq (4\pi/3) R^3\max\{ \|f_1\|_{L^\infty_x}, \|f_2\|_{L^\infty_x} \}\equiv K$, then by Cauchy-Schwarz' inequality and  Eq. (\ref{another wasserstein}), the integral $I_1$ can be estimated as
\begin{equation}\label{estimate I1}
I_1\leq C_K  \left\|\partial_xA_1-\partial_xA_2\right\|_{L^2_x} W_2(f_1,f_2).
\end{equation}
Similarly, using Cauchy-Schwarz' inequality and Lemma \ref{Lemma optimal transport}, we have that 
\begin{eqnarray*}
I_3  &\leq& 4\pi W_2(f_1,f_2) \int_1^2 \mbox{d}\theta\left(\int_{\R^3}\int_{\R^3} f_\theta |A_1-A_2|^2\mbox{d}x\,\mbox{d}p\right)^{1/2}\\
&\leq& 4\pi K \; W_2(f_1,f_2) \left(\int_{B_{2R}} |A_1-A_2|^2\,\mbox{d}x \right)^{1/2}
\end{eqnarray*}
and we deduce by Poincar\'e's inequality that 
\begin{equation}\label{estimate I3}
I_3 \leq  C_K  \left\|\partial_xA_1-\partial_xA_2\right\|_{L^2_x} W_2(f_1,f_2).
\end{equation}
As for $I_2$, Cauchy-Schwarz' inequality yields
\[I_2 \leq 4\pi W_2(f_1,f_2) \|\partial_xA_2\|^{1/2}_{L^\infty_x} \int_1^2 \mbox{d}\theta\left(\int_{\R^3}\int_{\R^3} f_\theta |A_1-A_2|^2\mbox{d}x\,\mbox{d}p\right)^{1/2}.\]
Then we use the second estimate in (\ref{LINFTY Derivative Potential}),  the estimate (\ref{f theta estimate}) of Lemma \ref{Lemma optimal transport}, and Poincar\'e's inequality to get, as in $I_3$,
\begin{equation}\label{estimate I2}
 I_2 \leq  C_K  \left\|\partial_xA_1-\partial_xA_2\right\|_{L^2_x} W_2(f_1,f_2).
\end{equation}
Combining (\ref{Estimate Three Integrals}) - (\ref{estimate I3}), we have 
\begin{equation}\label{equation simplified}
\left\|\partial_xA_1-\partial_xA_2\right\|^2_{L^2_x} + \left\|\rho_1^{1/2}(A_1-A_2)\right\|^2_{L^2_x} \leq C \left\|\partial_xA_1-\partial_xA_2\right\|_{L^2_x} W_2(f_1,f_2),
\end{equation}
for some constant $C>0$ that depends on $R$, $\left\|f_1\right\|_{L^{\infty}_{x,p}}$ and $\left\|f_2\right\|_{L^{\infty}_{x,p}}$. Since the left hand side of (\ref{equation simplified}) is bounded below by $\left\|\partial_xA_1-\partial_xA_2\right\|^2_{L^2_x}$, we deduce that
\begin{equation}\label{equation result 1}
\left\|\partial_xA_1-\partial_xA_2\right\|_{L^2_x} \leq C\, W_2(f_1, f_2).
\end{equation}
This proves the second estimate of (\ref{Potentials Estimate}). As for the estimate (\ref{Last Integral}), we insert (\ref{equation result 1}) in the right hand side of (\ref{equation simplified}), and we use the fact that the left hand side of (\ref{equation simplified}) is bounded below by $\left\|\rho_1^{1/2}(A_1-A_2)\right\|^2_{L^2_x}$ to obtain that
\[ \int_{\mathbb{R}^3}\rho_1(x)\left|A_1(x)-A_2(x)\right|^2dx\leq C \,W^2_2(f_1,f_2).\]
This completes the proof of the lemma. \qed


\section{Proof of Theorem \ref{Main Theorem}}\label{section proof}

For a non-negative function $f_0\in L^1\cap L^\infty(\R^3\times\R^3;\R^3)$ with compact support, let $f_1$ and $f_2$ be two compactly supported weak solutions of the RVD system with the same Cauchy datum $f_0$. Let $(\Phi_i,A_i)$ be the potentials induced by $f_i$, $i=1,2$, respectively. Denote $z=(x,p)$ and let $0\leq s\leq t< T$. To ease notation, write $Z_i(s,t)$ instead of $Z_i(s,t,z)$ for the solution of the characteristic system (\ref{Characteristics X})-(\ref{Characteristics P}) associated to the Vlasov equation (\ref{Vlasov In RVD}). Equivalently, $Z_i(s,t)$ with inverse $Z^{-1}_i(s,t)=Z_i(t,s)$ is the characteristic flow associated to the solution $f_i(t)=Z_i(t,s)_{\#}f_i(s)$ of the Vlasov equation; see  Lemma \ref{Definition Linear Vlasov} and Remark \ref{PushForwardRemark}. In particular $Z_i(t,0)$ is the flow associated to $f_i(t)=Z_i(t,0)_{\#}f_0$, that is, $f_i(t,z)=f_0(Z_i(0,t,z))$ for all $z\in B\subset\mathbb{R}^6$, $B$ a Borel set. We further denote $Z_i(t)$ instead of $Z_i(t,0)$ and define the function 
\begin{equation}
\label{Q}
Q(t)=\frac{1}{2}\int_{\mathbb{R}^6}f_0(z)\left|Z_1(t)-Z_2(t)\right|^2dz.
\end{equation}
We have $W^2_2(f_1(t),f_2(t))\leq 2Q(t)$ because $\gamma=\left(Z_1(t), Z_2(t)\right)_{\#}f_0$ is admissible in (\ref{def wasserstein}); here the function $\left(Z_1(t), Z_2(t)\right): \R^6 \rightarrow \R^6\times\R^6$ is defined by $\left(Z_1(t), Z_2(t)\right)(z) = \left(Z_1(t,z), Z_2(t,z)\right)$. Clearly, $Q(0)=0$. Our goal is to show that $Q(t)=0$ for every $t\in [0,T[$. If so, then $W_2(f_1(t),f_2(t))=0$ which implies that $f_1=f_2$ on $[0,T[\times\mathbb{R}^6$ and therefore uniqueness.  

Take the time derivative on both sides of (\ref{Q}). By Lemma \ref{Definition Linear Vlasov} we have
\begin{eqnarray}
\lefteqn{\dot{Q}(t) = \int_{\mathbb{R}^6}f_0(z)\left[\frac{}{}Z_1(t)-Z_2(t)\frac{}{}\right]\cdot\left[\dot{Z}_1(t)-\dot{Z}_2(t)\right]dz}\nonumber\\
& = &  \int_{\mathbb{R}^6}f_0(z)\left[\frac{}{}X_1(t)-X_2(t)\frac{}{}\right]\cdot\left[\frac{}{}v_{A_1}(t,Z_1(t))-v_{A_2}(t,Z_2(t))\frac{}{}\right]dz\nonumber\\
&  & -  \int_{\mathbb{R}^6}f_0(z)\left[\frac{}{}P_1(t)-P_2(t)\frac{}{}\right]\cdot\left[\frac{}{}\nabla\Phi_1(t,X_1(t))-\nabla\Phi_2(t,X_2(t))\frac{}{}\right]dz\nonumber\\
&  & +  \int_{\mathbb{R}^6}f_0(z)\left[\frac{}{}P_1(t)-P_2(t)\frac{}{}\right]\cdot\left[\frac{}{}v^i_{A_1}\nabla A^i_1(t,Z_1(t))-v^i_{A_2}\nabla A^i_2(t,Z_2(t))\frac{}{}\right]dz\nonumber\\
& =: & I_1(t)+I_2(t)+I_3(t).\nonumber 
\end{eqnarray}
In \cite{Loeper}, it is shown that for some constant $C>0$ that depends only on $\left\|\rho_i\right\|_{L^{\infty}_{t,x}}$,
\begin{equation}
\label{Estimate I2}
I_2(t)\leq CQ(t)\left(1-\ln Q(t)\right),
\end{equation}
provided $\left\|Z_1(t)-Z_2(t)\right\|_{L^{\infty}_z}\leq e^{-1}$. This is essentially the result in \cite{Loeper} which yields uniqueness of weak solutions of the Vlasov-Poisson system under the assumption that the charge density stays bounded. As for the RVD system, it remains to estimate $I_1$ and $I_3$. To estimate $I_1$, recall that $$v_A(t,X(t),P(t))=v(P(t)-A(t,X(t))).$$ where $v(g)=g\left(1+g^2\right)^{-1/2}$. 
Then, since $g\mapsto v(g)$ is $C^1_b$, we have
\begin{eqnarray*}
\lefteqn{\left|v_{A_1}(t,X_1(t),P_1(t))- v_{A_2}(t,X_2(t),P_2(t))\right|} \nonumber\\
& \leq &C\left( \left|P_1(t)-P_2(t)\right|+\left|A_1(t,X_1(t))-A_2(t,X_2(t))\right|\right)\nonumber\\
& \leq & C\left(\frac{}{}\left|P_1(t)-P_2(t)\right|+\left|X_1(t)-X_2(t)\right|+\left|A_1(t,X_1(t))-A_2(t,X_1(t))\right|\frac{}{}\right),
\end{eqnarray*}
where $C=C(\left\|\rho_i\right\|_{L^{\infty}_{t,x}})$. Note the use of the second estimate in (\ref{LINFTY Derivative Potential}) for the last step. Thus, Cauchy-Schwarz' inequality yields
\begin{equation}
\label{Estimate I1}
 I_1 \leq C\left(Q(t)+Q^{1/2}(t)T^{1/2}(t)\right),
\end{equation}
where, in view of (\ref{pushforward}), and (\ref{Last Integral}) in Lemma \ref{Potentials Estimate Lemma 1},
\begin{eqnarray}
\label{T1}
T(t) & = & \int_{\mathbb{R}^6}f_0(z)\left|A_1(t,X_1(t))-A_2(t,X_1(t))\right|^2dz\nonumber\\
&=& \int_{\mathbb{R}^3}\rho_0(x)\left|A_1(t,X_1(t))-A_2(t,X_1(t))\right|^2dx\nonumber\\
& = & \int_{\mathbb{R}^3}\rho_1(t,x)\left|A_1(t,x)-A_2(t,x)\right|^2dx\nonumber\nonumber\\
& \leq & CW^2_2(f_1,f_2),
\end{eqnarray}
with $C=C(R,\left\|f_i\right\|_{L^{\infty}_{t,x,p}})$. Therefore, since $W^2_2(f_1(t),f_2(t))\leq 2Q(t)$, we find that $I_1\leq CQ(t)$. 

The integral $I_3$ can be estimated just as $I_1$ and $I_2$. Indeed, we have
\begin{eqnarray}
\label{Working Estimate 2}
I_3 & = & \int_{\mathbb{R}^6}f_0(z)\left[\frac{}{}P_1(t)-P_2(t)\frac{}{}\right]\nonumber\\
& & \quad \cdot \left[\frac{}{}v^i_{A_1}(t,Z_1(t))-v^i_{A_2}(t,Z_2(t))\frac{}{}\right]\nabla A^i_1(t,X_1(t))dz\nonumber\\
& & +\int_{\mathbb{R}^6}f_0(z)\left[\frac{}{}P_1(t)-P_2(t)\frac{}{}\right]\nonumber\\
& & \qquad \cdot v^i_{A_2}(t,Z_2(t))\left[\frac{}{}\nabla A^i_1(t,X_1(t))-\nabla A^i_2(t,X_2(t))\frac{}{}\right]dz.\nonumber
\end{eqnarray}
In view of (\ref{LINFTY Derivative Potential}), the first integral on the right-hand side can be estimated exactly as $I_1$. On the other hand, the second integral on the right-hand side is analogous to $I_2$, with the vector potential instead of the scalar potential. Hence, since $\left|v_{A_2}\right|\leq 1$, we can use mutatis mutandi the arguments in \cite{Loeper} and Lemmas \ref{Estimate Darwin Potentials} and \ref{Potentials Estimate Lemma 1} to estimate this integral as in (\ref{Estimate I2}). 

Then, we gather all previous estimates to find that for some constant $C=C(R,\left\|f_i\right\|_{L^{\infty}_{t,x,p}})>0$
\begin{equation}
\dot{Q}(t) \leq CQ(t)\left(1-\ln Q(t)\right),
\end{equation}
whenever $\left\|Z_1(t)-Z_2(t)\right\|_{L^{\infty}_z}\leq e^{-1}$. This is a Gronwall's-type inequality which by standard arguments yields $Q(t)\equiv0$ on $[0,T[$ and therefore uniqueness. 
\qed


\section*{Acknowledgment} 
Martial Agueh is supported by a Discovery grant from the Natural Science and Engineering Research Council of Canada.


\end{document}